\documentclass[10pt,a4paper]{article}
\usepackage{a4wide}
\usepackage{graphicx}

\begin{document}
\renewcommand{\topfraction}{0.85}
\renewcommand{\bottomfraction}{0.7}
\renewcommand{\textfraction}{0.15}
\renewcommand{\floatpagefraction}{0.66}

\begin{center}
 {\Large \bf
Discovery of Very-High-Energy $\gamma$-Rays from the Galactic~Centre~Ridge}
\vspace{5mm}

F. Aharonian$^{1}$,
  A.G.~Akhperjanian$^{2}$,
  A.R.~Bazer-Bachi$^{3}$,
  M.~Beilicke$^{4}$,
  W.~Benbow$^{1}$,
  D.~Berge$^{1}$,
  K.~Bernl\"ohr$^{1,5}$,
  C.~Boisson$^{6}$,
  O.~Bolz$^{1}$,
  V.~Borrel$^{3}$
% ...
 
\vspace{4mm}\small $^{1}$ Max-Planck-Institut f\"ur Kernphysik, Heidelberg, Germany\\
\small $^{2}$ Yerevan Physics Institute, Armenia\\
\small $^{3}$ Centre d'Etude Spatiale des Rayonnements, CNRS/UPS, Toulouse, France\\
\small $^{4}$ Universit\"at Hamburg, Institut f\"ur Experimentalphysik, Germany\\
\small $^{5}$ Institut f\"ur Physik, Humboldt-Universit\"at zu Berlin, Germany\\
\small $^{6}$ LUTH, UMR 8102 du CNRS, Observatoire de Paris, Section
de Meudon, France\\
\small $^{7}$ University of Durham, Department of Physics, U.K.\\
\small $^{8}$ Laboratoire Leprince-Ringuet, IN2P3/CNRS,
 Ecole Polytechnique, Palaiseau, France\\
\small $^{9}$ APC (UMR 7164, CNRS, Universit\'e Paris VII, CEA, Observatoire de Paris), Paris\\ 
\small $^{10}$ Dublin Institute for Advanced Studies, Ireland\\
\small $^{11}$ Landessternwarte, K\"onigstuhl, D 69117 Heidelberg, Germany\\
\small $^{12}$ Laboratoire de Physique Th\'eorique et Astroparticules, IN2P3/CNRS,
 Universit\'e Montpellier II\\
\small $^{13}$ Laboratoire d'Astrophysique de Grenoble, INSU/CNRS, Universit\'e
 Joseph Fourier, France \\
\small $^{14}$ DAPNIA/DSM/CEA, CE Saclay, 
 Gif-sur-Yvette, France\\
\small $^{15}$ Unit for Space Physics, North-West University, Potchefstroom,
     South Africa \\
\small $^{16}$ Laboratoire de Physique Nucl\'eaire et de Hautes Energies, IN2P3/CNRS, Universit\'es
 Paris VI \& VII\\
\small $^{17}$ Institut f\"ur Theoretische Physik, Weltraum und
 Astrophysik, Ruhr-Universit\"at Bochum, Germany \\
\small $^{18}$ Institute of Particle and Nuclear Physics, Charles University,
     Prague, Czech Republic\\
\small $^{19}$ University of Namibia, Windhoek, Namibia\\
\small $^{20}$ European Associated Laboratory for Gamma-Ray Astronomy 

\end{center}



\noindent


\vspace{0.5cm}

{\bf 

The origin of Galactic cosmic rays (with energies up to $10^{15}$ eV)
remains unclear, though it is widely believed that they originate in
the shock waves of expanding supernova
remnants~\cite{CR_Review}\cite{HillasSNR}.  Currently the best way to
investigate their acceleration and propagation is by observing the
$\gamma$-rays produced when cosmic rays interact with interstellar
gas~\cite{AharonianClouds}. Here we report observations of an extended
region of very high energy (VHE, $>$100 GeV) $\gamma$-ray emission
correlated spatially with a complex of giant molecular clouds in the
central 200 pc of the Milky Way.  The hardness of the $\gamma$-ray
spectrum and the conditions in those molecular clouds indicate that
the cosmic rays giving rise to the $\gamma$-rays are likely to be protons
and nuclei rather than electrons.  The energy associated with the
cosmic rays could have come from a single supernova explosion around
$10^{4}$ years ago.

}\pagebreak

The observations described here were carried out with the High Energy
Stereoscopic System (H.E.S.S.), a system of four imaging
atmospheric-Cherenkov telescopes \cite{HESS}. The instrument operates
in the Teraelectronvolt energy range (TeV), beyond the regime
accessible to satellite-based detectors (MeV up to $\sim$10~GeV).  At
satellite energies, the technique of probing the distribution of
cosmic rays (CRs) using $\gamma$-ray emission has been demonstrated in
the large-scale mapping of the Galactic plane by
EGRET~\cite{EGRET_DIFFUSE}.  The $\gamma$-ray flux was found to
approximately trace the density of interstellar gas, illustrating that
the flux of CRs is roughly constant throughout the Galaxy.
However, given its modest angular resolution ($\sim1^{\circ}$), EGRET
could only resolve the few nearest molecular clouds.  The order of
magnitude better angular resolution of H.E.S.S.  opens up this
possibility of resolving individual clouds out to the distance of the
Galactic Centre (GC). Moreover, in the energy range accessible to
EGRET, the picture is complicated by the contribution of cosmic
electrons~\cite{CR_Review} to the diffuse $\gamma$-ray flux via
inverse Compton (IC) scattering and Bremsstrahlung.  In the energy range of
H.E.S.S. the dominant component of the truely diffuse $\gamma$-ray
emission is very likely the decay of neutral pions produced in the
interactions of CRs with ambient material.  Taken together, the wide
field of view ($\sim$5$^{\circ}$) and the improved angular resolution
(better than 0.1$^{\circ}$) of H.E.S.S. have made possible the mapping
of extended $\gamma$-ray emission.

\begin{figure}
  \begin{center}
    \includegraphics[width=14cm]{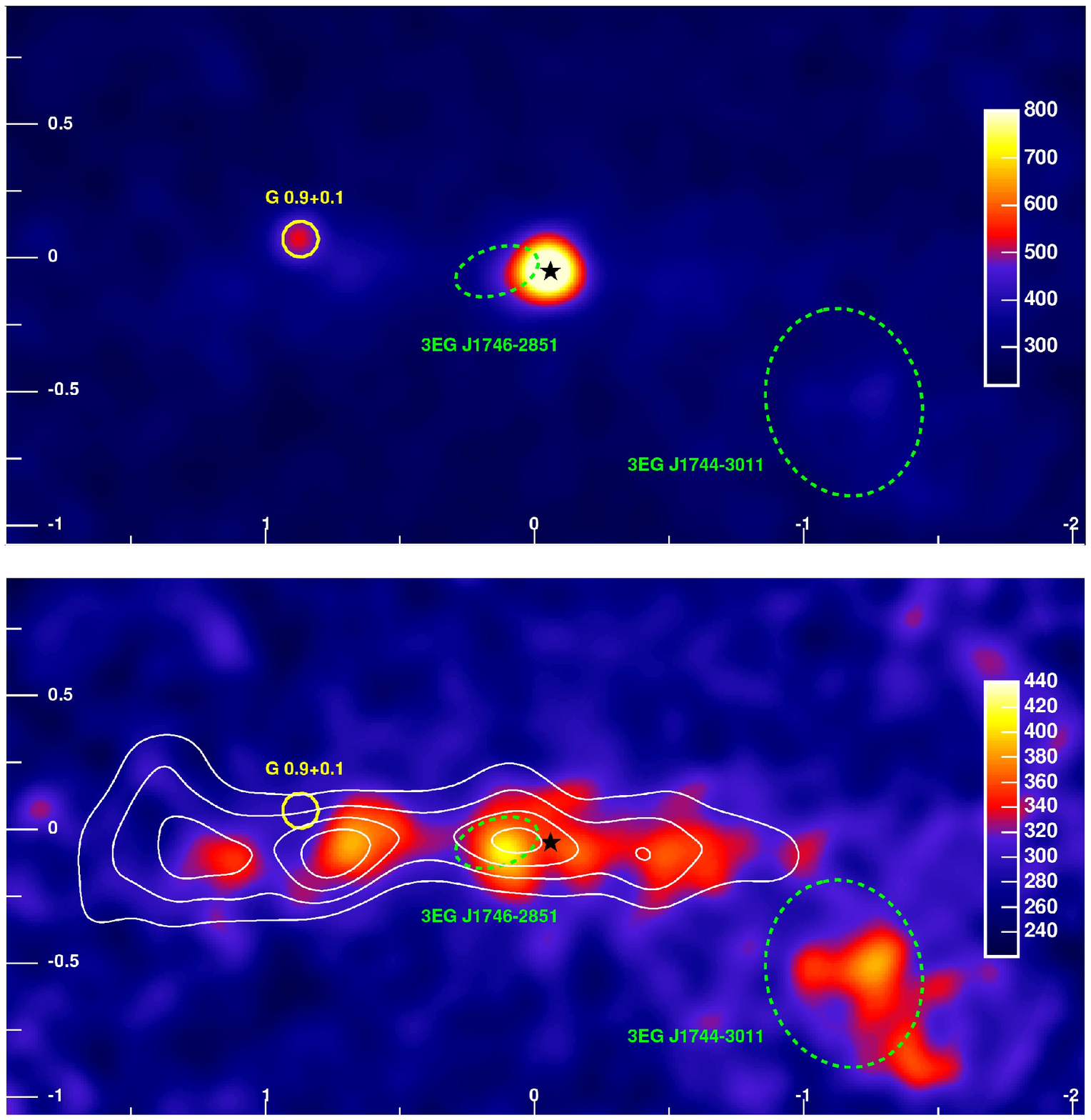}
  \end{center}
  \caption{VHE $\gamma$-ray images of the GC region. Top:
$\gamma$-ray count map, bottom: the same map after subtraction of the
two dominant point sources, showing an extended band of gamma-ray
emission. White contour lines indicate the density of molecular gas,
traced by its CS emission. The position and size of the composite SNR
G\,0.9+0.1 is shown with a yellow circle. The position of
Sgr~A$^{\star}$ is marked with a black star.  The 95\% confidence
region for the positions of the two unidentified EGRET sources in the
region are shown as dashed green ellipses~\cite{EGRETEllipses}.  These
smoothed and acceptance corrected images are derived from 55 hours of
data consisting of dedicated observations of Sgr\,A$^{\star}$,
G\,0.9+0.1 and a part of the data of the H.E.S.S. Galactic plane
survey~\cite{HESSSCAN}. The excess observed along the Galactic plane
consists of $\approx$3500 $\gamma$-ray photons and has a statistical
significance of 14.6 standard deviations.  The absence of any residual
emission at the position of the point-like $\gamma$-ray source
G\,0.9+0.1 demonstrates the validity of the subtraction technique. The
energy threshold of the maps is 380~GeV due to the tight
$\gamma$-ray selection cuts applied here to improve signal/noise and
angular resolution. We note that the ability of H.E.S.S. to map
extended $\gamma$-ray emission has been demonstrated for the
shell-type SNRs RX\,J1713.7 $-$3946~\cite{HESS_RXJ1713} and
RX\,J0852.0$-$4622~\cite{HESS_VelaJnr}.  The white contours are evenly
spaced and show velocity integrated CS line emission from
Tsuboi~et~al.~\cite{Tsuboi}, and have been smoothed to match the
angular resolution of H.E.S.S..
  }
  \label{fig_map}
\end{figure}

\begin{figure}
  \begin{center}
    \includegraphics[width=13cm]{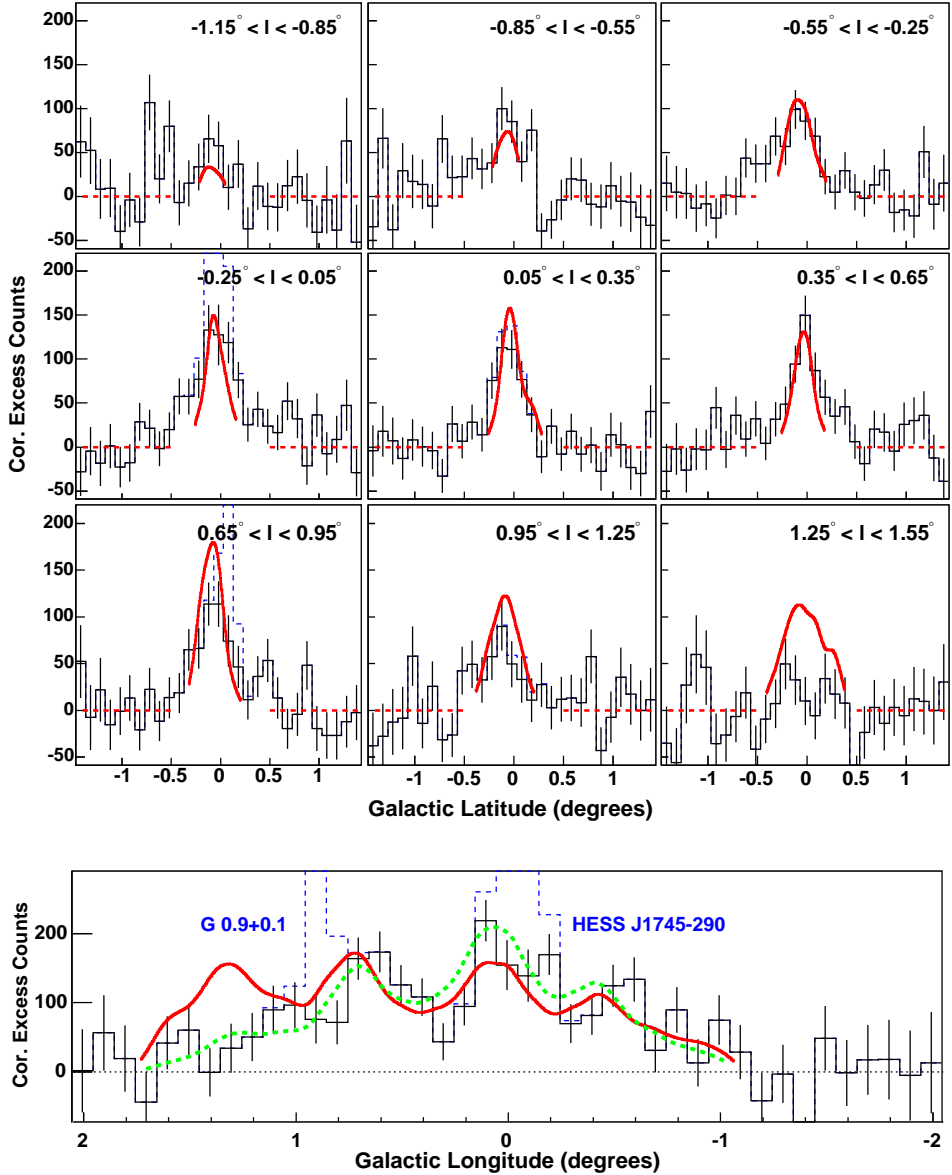}
  \end{center}
  \caption{Distribution of $\gamma$-ray emission in Galactic
latitude (for individual slices in longitude, top) and in Galactic
longitude (bottom). The red curves show the density of molecular gas,
traced by CS emission. The upper curves show acceptance corrected (and
cosmic-ray background subtracted) $\gamma$-ray counts for
0.3$^{\circ}$ wide bands in longitude. The point-source subtracted
counts are shown in black. The dashed blue histogram shows the
unsubtracted values (the $y$-scale is truncated). The red curves
correspond to the smoothed CS map of Fig.~\ref{fig_map} and are drawn
only in the regions where CS measurements are available. The dashed
red lines show nominal zero CS density in regions away from the
Galactic plane.  The lower plot shows $\gamma$-ray counts versus $l$
for $-0.2^{\circ}<b<0.2^{\circ}$.  The CS line flux may be
underestimated close to $l=-1^{\circ}$ due to a narrower coverage in
$b$ at this longitude. The dashed line shows the $\gamma$-ray flux
expected if the CR density distribution can be described by a Gaussian
centred at $l=0^{\circ}$ and with rms $0.8^{\circ}$, as expected in a
simple model for diffusion away from a central source of age $\sim
10^{4}$ years.  In all plots the background level is estimated using
events from the regions $0.8^{\circ}<|b|<1.5^{\circ}$. Error bars show
$\pm1$ standard deviation.
  }
  \label{fig_slices}
\end{figure}

\begin{figure}
  \begin{center}
    \includegraphics[width=13cm]{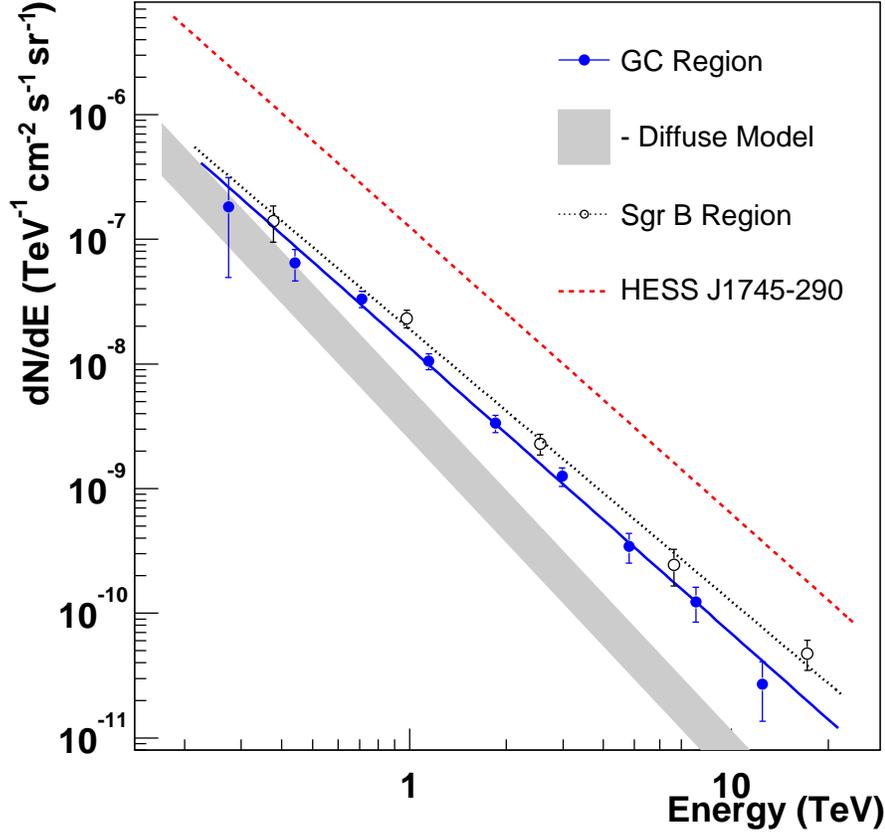}
  \end{center}
  \caption{
$\gamma$-ray flux per unit solid angle in the GC
region (data points), in comparison with the expected flux assuming 
a cosmic-ray spectrum as measured in the solar neighbourhood (shaded
band). The spectrum of the region $-0.8^{\circ}<l<0.8^{\circ}$,
$|b|<0.3^{\circ}$ is shown using full circles. These data can be
described by a power law: $dN/dE = k(E/\mathrm{TeV})^{-\Gamma}$, with 
$k = (1.73 \pm0.13_{stat}\pm0.35_{sys}) \times
10^{-8}$ TeV$^{-1}$ cm$^{-2}$ s$^{-1}$ sr$^{-1}$ and a photon index
$\Gamma=2.29\pm0.07_{stat}\pm0.20_{sys}$.  The shaded box shows the
range of expected $\pi^{0}$-decay fluxes from this region assuming a
CR spectrum identical to that found in the solar neighbourhood and a
total mass of $1.7-4.4\times10^{7}$ solar masses in the region
$-0.8^{\circ}<l<0.8^{\circ}$, $|b|<0.3^{\circ}$, estimated from CS
measurements. Above 1~TeV an enhancement by a factor $3-9$ relative to
this prediction is observed.  Using independent mass estimates derived
from sub-millimeter measurements~\cite{SCUBA}, $5.3\pm1.0\times10^{7}$
solar masses, and from C$^{18}$O measurements~\cite{C18O},
$3^{+2}_{-1}\times10^{7}$ solar masses, results in enhancement factors
of $4-6$ and $5-13$, respectively.
The strongest emission away from the bright central source
HESS\,J1745$-$290 occurs close to the Sagittarius B complex of giant
molecular clouds~\cite{Lis}. In a box covering this region
($0.3^{\circ}<l<0.8^{\circ}$, $-0.3^{\circ}<b<0.2^{\circ}$),
integrated CS emission suggests a molecular target mass of
$6-15\times10^{6}$ solar masses. The energy spectrum of this region is
shown using open circles. The measured $\gamma$-ray flux ($>$ 1~TeV)
implies a high-energy cosmic-ray density which is $4-10$ times higher
than the local value. Standard $\gamma$-ray selection cuts are applied
here, yielding a spectral analysis threshold of 170~GeV. The spectrum
of the central source HESS\,J1745-290 is shown for comparison (using
an integration radius of $0.14^{\circ}$). All error bars
show $\pm1$ standard deviation.
}
  \label{fig_spectrum}
\end{figure}

Early H.E.S.S. observations of the GC region led to the detection of a
point-like source of VHE $\gamma$-rays at the gravitational centre of
the Galaxy (HESS J1745$-$290)~\cite{HESSSAGA}, compatible with the
positions of the supermassive black hole Sagittarius~A$^{\star}$, the
supernova remnant (SNR) Sgr~A~East, and a GC source reported by other
groups~\cite{CANGAROO,VERITAS}. A more sensitive exposure of the
region in 2004 revealed a second source: the supernova remnant/pulsar
wind nebula G\,0.9+0.1~\cite{HESSG09}.  These two sources are clearly
visible in the upper panel of Fig.~\ref{fig_map}. For previous VHE
instruments such sources were close to the limit of
detectability. With the greater sensitivity of the H.E.S.S. instrument
it is possible to subtract these two sources and search for much
fainter emission. Subtracting the best fit model for point-like
emission at the position of these excesses yields the map shown in
Fig.~\ref{fig_map} (bottom). Two significant features are apparent
after subtraction: extended emission spatially coincident with the
unidentified EGRET source 3EG\,J1744-3011 (discussed
elsewhere~\cite{HESSSCAN2}) and emission extending along the Galactic
plane for roughly 2$^{\circ}$.  The latter emission is not only very
clearly extended in longitude $l$, but also significantly extended in
latitude $b$ (beyond the angular resolution of H.E.S.S.) with a
characteristic root mean square (rms) width of 0.2$^{\circ}$, as can be seen
in the Galactic latitude slices shown in Fig.~\ref{fig_slices}. The
reconstructed $\gamma$-ray spectrum for the region
$-0.8^{\circ}<l<0.8^{\circ}$, $|b|< 0.3^{\circ}$ (with point-source
emission subtracted) is well described by a power law with photon
index $\Gamma = 2.29\pm0.07_{stat}\pm0.20_{sys}$
(Fig.~\ref{fig_spectrum}).

Given the plausible assumption that the $\gamma$-ray emission takes
place near the centre of the Galaxy, at a distance of about 8.5 kpc,
the observed rms extension in latitude of $0.2^{\circ}$ corresponds to
a scale of $\approx$ 30~pc. This value is similar to that of
interstellar material in giant molecular clouds in this region, as
traced by their CO emission and in particular by their CS emission
\cite{Tsuboi}. CS line emission does not suffer from the problem of
`standard' CO lines \cite{Oka}, that clouds are optically thick for
these lines and hence the total mass of clouds may be underestimated.
The CS data suggest that the central region of the Galaxy, $|l| <
1.5^{\circ}$ and $|b| < 0.25^{\circ}$, contains about
$3-8\times10^{7}$ solar masses of interstellar gas, structured in a
number of overlapping clouds, which provide an efficient target for
the nucleonic CRs permeating these clouds.  The region over
which the $\gamma$-ray spectrum is integrated contains 55\% of the CS
emission, corresponding to a mass of $1.7-4.4\times10^{7}$ solar
masses. At least for $|l|<1^{\circ}$, we find a close match between
the distribution of the VHE $\gamma$-ray emission and the density of
dense interstellar gas as traced by CS emission (Fig.~\ref{fig_map}
(bottom) and Fig.~\ref{fig_slices}).

The close correlation between $\gamma$-ray emission and available
target material in the central 200 pc of our galaxy is a strong
indication for an origin of this emission in the interactions of
CRs. Following this interpretation, the similarity in the
distributions of CS line and VHE $\gamma$-ray emission implies a
rather uniform CR density in the region.  Since in the case of a
power-law energy distribution the spectral index of the $\gamma$-rays
closely traces the spectral index of the CRs themselves (corrections
due to scaling violations in the CR interactions are small,
$\Delta\Gamma\,<\,0.1$), the measured
$\gamma$-ray spectrum implies a CR spectrum near the
GC with a spectral index close to 2.3, significantly
harder than in the solar neighbourhood (where an index of 2.75 is
measured). Given the probable proximity of particle accelerators,
propagation effects are likely to be less pronounced than in the
Galaxy as a whole, providing a natural explanation for the harder
spectrum which is closer to the intrinsic CR-source spectra.  The main
uncertainty in estimating the flux of CRs in the GC is
the uncertainty in the amount of target material. Following
\cite{AharonianClouds} and using the mass estimate of
Tsuboi~\cite{Tsuboi} we can estimate the expected $\gamma$-ray flux
from the region, assuming for the moment that the GC cosmic-ray flux
and spectrum are identical to those measured in the solar
neighbourhood. Fig.~\ref{fig_spectrum} shows the expected $\gamma$-ray
flux as a grey band, together with the observed spectrum. Whilst below
500~GeV there is reasonable agreement with this simple prediction,
there is a clear excess of high energy $\gamma$-rays over
expectations.  The $\gamma$-ray flux above 1~TeV is a factor $3-9$
higher than the expected flux.  The implication is that the number
density of CRs with multi-TeV energies exceeds the local
density by the same factor. The size of the enhancement
increases rapidly at energies above 1~TeV.

The observation of correlation between target material and TeV
$\gamma$-ray emission is unique and provides a compelling case for an
origin of the emission in the interactions of CR nuclei.  In addition,
the key experimental facts of a harder than expected spectrum, and a
higher than expected TeV flux, imply that there is an additional
component to the GC cosmic-ray population above the CR `sea' which
fills the Galaxy. This is the first time that such direct evidence for
recently accelerated (hadronic) CRs in any part of our galaxy has been
found. The energy required to accelerate this additional component is
estimated to be $10^{49}$ erg in the energy range 4-40 TeV or 
$\sim\,10^{50}$ erg in total if the measured spectrum extends from
$10^{9}-10^{15}$ eV. Given a typical supernova explosion energy of
$10^{51}$ erg, the observed CR excess could have been produced in a
single SNR, assuming a 10\% efficiency for CR acceleration. Following
such a scenario, any epoch of CR production must have occurred in the
recent enough past that the CRs accelerated have not yet diffused out
of the GC region.  Representing the diffusion of protons with energies
of several TeV in the form $D=\eta 10^{30} \ \rm cm^{2} s^{-1}$, where
$10^{30} \ \rm cm^{2} s^{-1}$ is the approximate value of the
diffusion coefficient in the Galactic Disk at TeV energies, we
estimate the diffusion time-scale to be $t = R^{2}/2D \approx 3000
(\theta\,/\,1^{\circ})^{2}/\eta$ years, where $\theta$ is the angular
distance from the GC. Due to the larger magnetic field and higher
turbulence in the central region compared to more conventional regions
of the Galactic disk, the normalisation parameter $\eta$ is likely
$\leq 1$ and a source or sources of age $\sim$10~kyr could fill the
region $|l|<1^{\circ}$ with CRs. Indeed, the observation of a deficit
in VHE emission at $l=1.3^{\circ}$ relative to the available target
material (see Fig.~\ref{fig_slices}) suggests that CRs, which were
recently accelerated in a source or sources in the GC region, have not
yet diffused out beyond $|l|=1^{\circ}$.

The observed morphology and spectrum of the $\gamma$-ray emission 
provide evidence that one or more cosmic-ray accelerators have been active in
the GC in the last 10,000 years. The fact that the diffuse emission
exhibits a photon index $\Gamma$ which is the same - within errors -
as that of the central source HESS\,J1745$-$290 suggests that this
object could be the source in question. Within the 1 arcminute error
box of HESS\,J1745$-$290 are two compelling candidates for such a CR
accelerator. The first is the SNR Sgr A East~\cite{Melia} with its
estimated age around 10~kyr~\cite{ChandraAEast} (younger ages have
been quoted for Sgr A East~\cite{AEastAge} reflecting the significant
uncertainty in this estimate). The second is the supermassive black
hole Sgr~A$^{\star}$~\cite{AharonianSAGA, Belanger} which may have
been more active in the past.

A distinct alternative possibility is that a population of 
\emph{electron} accelerators produces the observed $\gamma$-ray
emission via IC scattering. Extended objects with
photon indices close to the value $2.3$ observed in the GC are
observed elsewhere in the Galactic plane~\cite{HESSSCAN2}.  The parent
population of objects such as pulsar wind nebulae (i.e. massive stars)
would likely follow approximately the distribution of molecular gas.
However, in the intense photon fields and high magnetic fields within
and close to the GC molecular clouds~\cite{BFIELD,GCReview}, TeV
electrons would lose their energy rapidly: $t_{\mathrm{rad}} \approx
120\,(B/100\,\mu\mathrm{G})^{-2}\,(E_{e}/10\,\mathrm{TeV})^{-1}$
years. We would therefore expect to see rather compact sources
(point-like for H.E.S.S.) which would also be bright in the X-ray
regime (as is for example G\,0.9+0.1). The existence of $\sim$ 10 such
unknown sources in this small region again seems unlikely. Any
substantially extended IC source would most likely be a foreground
source along the line-of-sight towards the GC region, making any
correlation with GC molecular clouds entirely coincidental.

\section*{Acknowledgements}

The support of the Namibian authorities and of the University of
Namibia in facilitating the construction and operation of H.E.S.S. is
gratefully acknowledged, as is the support by the German Ministry for
Education and Research (BMBF), the Max Planck Society, the French
Ministry for Research, the CNRS-IN2P3 and the Astroparticle
Interdisciplinary Programme of the CNRS, the U.K. Particle Physics and
Astronomy Research Council (PPARC), the IPNP of the Charles
University, the South African Department of Science and Technology and
National Research Foundation, and by the University of Namibia. We
would like to thank M. Tsuboi for providing the CS survey data used
here and Y. Moriguchi and Y. Fukui for helpful discussions on
molecular tracers.

{\bf Correspondance} should be addressed to J.A. Hinton, \emph{\bf Jim.Hinton@mpi-hd.mpg.de}

\end{document}